\documentclass[structabstract]{aa}  
\usepackage{graphicx}
\usepackage{txfonts}
\usepackage{longtable}

\begin{document}
   \title{The Horizontal Branch luminosity vs metallicity in M31 
   globular clusters.\thanks{Based on observations made with the NASA/ESA {\em Hubble 
   Space Telescope}, obtained from the Hubble Legacy Archive, which is a collaboration 
   between the Space Telescope Science Institute (STScI/NASA), the Space Telescope 
   European Coordinating Facility (ST-ECF/ESA) and the Canadian Astronomy Data 
   Centre (CADC/NRC/CSA). STScI is operated by the Association of Universities for 
   Research in Astronomy, Inc., under NASA contract NAS 5-26555.}, \thanks{Photometric catalogs are available
    at the CDS via anonymous ftp to cdsarc.u-strasbg.fr (130.79.128.5) or via http://cdsarc.u-strasbg.fr/viz-bin/qcat?J/A+A/
     and at http://www.bo.astro.it/M31/hstcatalog/}}
   \titlerunning{V(HB) vs [Fe/H] in M31 Globular Clusters}

   \author{L. Federici\inst{}, C. Cacciari\inst{}, M. Bellazzini\inst{}, F. Fusi Pecci\inst{}, S. Galleti\inst{}            
          \and
          S. Perina\inst{}
          }
	  
   \authorrunning{Federici et al.} 
   \institute{INAF Osservatorio Astronomico di Bologna, 
              Via Ranzani 1, Bologna, 40127-I, Italy \\
              \email{[luciana.federici, carla.cacciari, michele.bellazzini,  
	      flavio.fusipecci, silvia.galleti, sibilla.perina] at oabo.inaf.it }
             }
   \date{Received 01/04/2012; accepted 12/07/2012}

 
  \abstract
   {Thanks to the outstanding capabilites of the $HST$, our current knowledge about the M31 
   globular clusters (GCs) is similar to our knowledge of the Milky Way GCs in the 1960s-1970s,  
   which set the basis for studying the halo and galaxy formation using these objects as tracers, 
   and established their importance in defining the cosmic distance scale.} 
   {We intend to derive a new calibration of the M$_V$(HB)-[Fe/H] relation by exploiting 
   the  large photometric database of old GCs in M31 in the $HST$ archive.}
   {We collected the BVI data for 48 old GCs in M31 and analysed them by applying the same 
   methods and procedures to all objects. We obtained a set of homogeneous colour-magnitude 
   diagrams (CMDs) that were best-fitted with the fiducial CMD ridge lines of 
   selected Milky Way 
   template GCs. Reddening, metallicity, Horizontal Branch (HB) luminosity and distance were 
   determined self-consistently for each cluster.}  
   {There are three main results of this study:  
   i)  the relation M$_V$(HB)=0.25($\pm$0.02)[Fe/H]+0.89($\pm$0.03), which is 
   obtained from the above parameters and is calibrated on the distances of the template
   Galactic GCs;
   ii) the distance modulus to M31 of (m-M)$_0$=24.42$\pm$0.06 mag, obtained 
   by normalising this relation at the reference value of [Fe/H]=--1.5 to a 
   similar relation using  V$_0$(HB). This is the first determination of the distance 
   to M31 based on the characteristics of its GC system which is calibrated 
   on Galactic GCs;
   iii) the distance to the Large Magellanic Cloud (LMC), which is estimated to be 
   18.54$\pm$0.07 mag 
   as a consequence of the previous results. These values agree excellently 
   with the most recent estimate based on $HST$ parallaxes of Galactic Cepheid 
   and RR Lyrae stars, as well as with recent methods.  
   } 
   {}
   \maketitle
   
   
%

\section{Introduction}

The globular cluster (GC) system of a galaxy is an important tracer of 
 its oldest stellar component, and hence gives information on the 
formation and evolution process primarily of the halo, and then of the galaxy 
as a whole. 
The systematic study of the Milky Way (MW) GCs, which had started in the '50s,  
produced colour-magnitude diagrams (CMD) and metal abundances for 19 GCs 
that led Searle and Zinn (1978) to challenge the Galaxy formation model proposed 
by Eggen, Lynden-Bell and Sandage (1962), namely the rapid collapse of a 
primordial gas cloud probably some 10 billion yr ago. 
Instead, Searle and Zinn (1978) proposed an accretion model for the Galactic halo of "transient 
protogalactic fragments that continued to fall into dynamical equilibrium with 
the Galaxy for some time after the collapse of its central regions had been 
completed." 
The classic works by Morgan (1959) and Kinman (1959) showed that 
there are two distinct populations of GCs in the Galaxy. 
The properties of these two populations were derived 
by Zinn (1985), who continued and further refined his previous analysis with the 
addition of CMDs, kinematics and metallicities for $\sim$ 120 GC, and  
showed that they have a very heterogeneous structure, kinematics and metallicities: 
the halo population is metal poor ([Fe/H] $<$ --0.8) and slowly
rotating with a roughly spherical distribution; the disk population is 
metal rich ([Fe/H] $>$ --0.8) and in rapid rotation. 
The past 20 years of $HST$ observations have made a dramatic contribution to 
our knowledge not only of the MW GCs, but also of the GC systems in other 
galaxies  (see Freeman and Bland-Hawthorn 2002 for a review of the first 10 
yr of $HST$ results), and the scenario is now much more detailed and complex. 

In this paper we focus our attention on another characteristic 
of  GCs, namely the very important role that they play for of the cosmic 
distance scale definition, because they host `standard candles'  such as RR Lyrae variables 
(Benedict et al. 2011; Caputo 2012), white dwarfs (Renzini et al. 1996; Hansen et al. 2007), 
and red giant stars at the very tip of the Red Giant Branch (TRGB) phase (Salaris 2012). 
Distances were derived  for a dozen GCs by fitting their main-sequence with local 
subdwarfs of known parallaxes (Gratton et al. 2003), and in a few cases by using 
member eclipsing binaries (Thompson et al. 2010). 
In addition, the GC system as a whole can be regarded as a standard candle 
for early-type giant galaxies, because the integrated luminosity function 
of the metal-poor GC subpopulation peaks at a nearly 
constant value of M$_V$ = --7.66$\pm$0.09 mag (Brodie \& Strader 2006; 
Rejkuba 2012). 

The aim of the present study is to exploit the large photometric database of M31 
GCs in the $HST$ archive and use the horizontal branch (HB) luminosity level V(HB) 
of their individual CMDs to derive a new calibration of the M$_V$(HB)-[Fe/H] 
relation by comparison with a reference grid of Galactic GCs. 
Despite of the great progress on distance determinations made in the last decade, 
there are still significant discrepancies among the results from various methods,  
also because our requirements have become more stringent in the meantime. 
Taking for example the Large Magellanic Cloud (LMC) as a reference 
place for comparing multiple distance 
indicators for consistency, we see that the individual distances span a range of 
about 0.10-0.20 mag, depending on the method (Walker 2011).
Most of the discrepancies are now due to systematic/calibration effects, and for 
this reason it is very important to provide a new calibration for such a widely 
used distance determination tool, based on the established ground of the MW GC 
system.    
 
The $HST$ archive presently contains multiband photometric data for 52 old
GCs in M31, which  can be used to obtain CMDs. 
These CMDs are still not deep enough to reach the main-sequence turn off (TO) 
and allow a direct age determination,  with the exception of cluster 
B379, which was observed for 120 $HST$ orbits and reached about 
1.5 mag fainter than the  TO (Brown et al. 2004). 
However, for 48 of these GCs the upper parts of the CMD, namely the  HB and the 
red giant branch (RGB), are clear and well defined, and are quite adequate for 
estimating  important parameters such as metallicity, reddening and 
distance.\footnote{The four clusters not considered in this study lie in the 
bulge region. They are B109, B115 and B143, whose photometry does not reach 
the HB (Jablonka et al. 2000), and B112 which was observed only in the JK bands 
(Stephens et al. 2001).}   
Therefore, the present situation for the GCs in 
M31 is not much different from the situation of the MW GCs in the 
1960s-1970s that set the basis for studying the halo and galaxy formation 
using these tracers. 

We have collected the BVI data for the 48 suitable M31 GCs and analysed them by
applying the same methods and procedures throughout to obtain a set of homogeneous CMDs, from 
which we derive reddening, metallicity, distance, as well as luminosity level of the HB.  
Much information on reddening and metallicity is 
available in the literature, but there are large discrepancies between the studies 
because of different procedures, assumptions and approximations as well as
observational errors. It is very important that these parameters are derived 
in a consistent and comparable way to minimise at least the systematics caused by 
different data treatment. From these parameters we derive the M$_V$(HB)-[Fe/H] 
relation defined by the largest and most accurate sample of M31 GCs so far, and 
also the first determination of the distance to M31 based on the characteristics of
its GC system calibrated on Galactic GCs. 

This study is the continuation of a long-term programme on 
the M31 GC system that was started more than two decades ago by our group, which  
focused on  the search for GC candidates (leading to the Revised Bologna 
Catalogue by Galleti et al. 2004, and web update\footnote{http://www.bo.astro.it/M31/}, 
hereafter RBCV4.0), and on the analysis of the properties of as many individual GCs 
as possible using the $HST$ outstanding imaging capabilities (Fusi-Pecci et al. 1996, 
hereafter FFP96; Rich et al. 2005, hereafter R05; Galleti et al. 2006, hereafter G06; 
Perina et al. 2009, hereafter P09; Perina et al. 2011, hereafter P11).    

The data are presented in Sect. \ref{s:data}, the method to derive reddening, 
metallicity, HB luminosity level and distance is discussed and applied in 
Sect.\ref{s:redmetlum}, 
the HB luminosity vs metallicity relation and its implication  for the distance estimate 
are discussed in Sect. \ref{s:hbmet}, a discussion on systematics is presented in 
Sect. \ref{s:syst}, and the summary and conclusions are given in Sect. \ref{s:sum}.


\section{The data}\label{s:data}

\subsection{The targets}

\begin{figure}
\centering
\includegraphics[width=9.0cm]{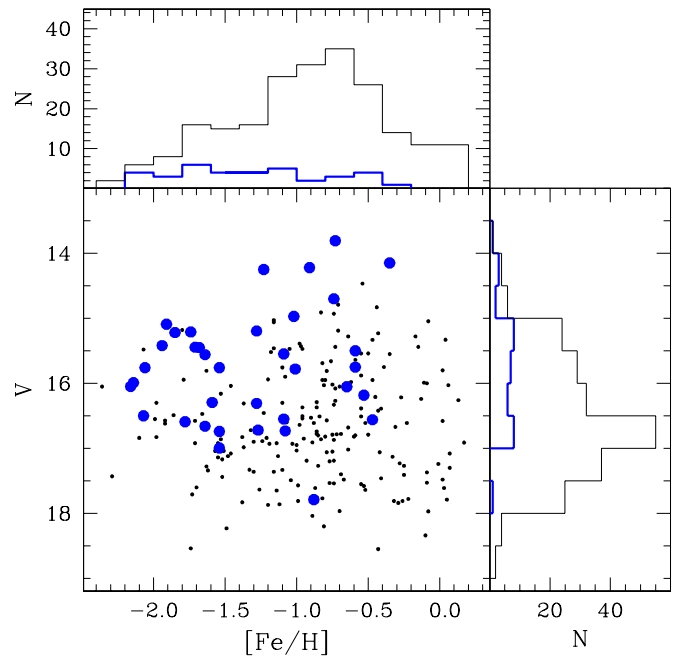}
\caption{Distribution of our cluster sample (shown by large blue dots and  
blue histogram contours) in the integrated V magnitude and metallicity space is compared
to the whole catalogue of confirmed M31 GCs (shown by small black dots and 
black histogram contours). The V magnitude and metallicity values are taken from 
RBCV4.0.
}
\label{f:vfehist}
\end{figure}

The present sample of 48 GCs, listed in Table \ref{t:targets}, corresponds to 
about 7\% of the total currently confirmed GC population in M31 (see RBCV4.0). 
They were originally selected for observation 
according to different purposes and criteria, e.g. brightness, colour, 
metallicity, and position. Therefore, the sample is somewhat biased towards brighter 
sources, except for those few faint objects that happened to fall into parallel fields. 
Moreover, the spatial sampling is not uniform, because targets were originally selected 
either to avoid  crowding or because they lay in particularly interesting areas. 
On the other hand, the metallicity distribution of our targets is likely to 
be well-representative of the entire GC population, because  several studies 
tried to sample the metallicity range as well as possible. 
The distribution of our sample is compared to the distribution of the whole catalogue 
of confirmed M31 GCs 
in the magnitude-metallicity space in Fig. \ref{f:vfehist}. The 
values of integrated V cluster magnitude and metallicity (from narrow-band 
spectrophotometry) were taken from RBCV4.0. 

However, none of these possible selection effects should affect our analysis in any 
significant way.

\begin{table*}[t]
\caption{Target GCs in M31.  Identification, integrated photometry 
and galactocentric projected coordinates X, Y, R (in arcmin)  are from the RBCV4.0. 
The second column (Ref.) refers to the papers presenting the original CMDs. The values of reddening, 
metallicity, V(HB) magnitude and distance modulus in columns 7-10 are the results of the present 
analysis (in particular, the V(HB) values include the correction of 0.08 mag for the clusters 
at [Fe/H]$>$ -1.0 - see text).
\label{t:targets}}
\begin{center}
\begin{tabular}[t]{lccrrrcccc}
\hline\noalign{\smallskip}
GC ID   & Ref. & V &\multicolumn{1}{c}{X} &\multicolumn{1}{c}{Y}&\multicolumn{1}{c}{R} & E(B--V) & [Fe/H] &   V(HB) & (m-M)$_{0}$ \\  
        &      &   &(arcmin)&(arcmin)&(arcmin)&         &        &         &             \\   
\noalign{\smallskip}
\hline\noalign{\smallskip}
				       
   B006-G058 & 	R05 & 15.50  &  -6.94$\;\;$ &  27.35$\;\;$ &   28.22$\;\;$ &  0.08 &  -0.55 & 25.46  & 24.56 \\  
   B008-G060 &	P09 & 16.56  & -15.45$\;\;$ &  19.89$\;\;$ &   25.18$\;\;$ &  0.07 &  -1.00 & 25.26  & 24.45 \\  
   B010-G062 &	P09 & 16.66  & -16.70$\;\;$ &  18.62$\;\;$ &   25.01$\;\;$ &  0.16 &  -1.80 & 25.28  & 24.30 \\  
   B012-G064 &	R05 & 15.09  & -10.77$\;\;$ &  22.98$\;\;$ &   25.38$\;\;$ &  0.11 &  -1.80 & 25.05  & 24.31 \\  
   B023-G078 &	P09 & 14.22  & -13.79$\;\;$ &  13.83$\;\;$ &   19.53$\;\;$ &  0.28: &  -0.90: & 25.91:  & 24.26: \\  
   B027-G087 &	R05 & 15.56  & -26.42$\;\;$ &   0.88$\;\;$ &   26.43$\;\;$ &  0.18: &  -1.66: & 25.53:  & 24.52: \\  
   B045-G108 &  R05 & 15.78  &   7.29$\;\;$ &  20.22$\;\;$ &   21.50$\;\;$ &  0.16 &  -0.90 & 25.62  & 24.55 \\  
   B058-G119 &	P11 & 14.97  & -28.82$\;\;$ & -10.19$\;\;$ &   30.57$\;\;$ &  0.11 &  -1.40 & 25.25  & 24.35 \\  
   B088-G150 &	P09 & 15.42  &   9.99$\;\;$ &  13.33$\;\;$ &   16.66$\;\;$ &  0.38 &  -1.90 & 25.99  & 24.44 \\  
   B158-G213 &	P09 & 14.70  &  -3.44$\;\;$ &  -9.88$\;\;$ &   10.47$\;\;$ &  0.09: &  -0.90: & 25.44:  & 24.42: \\  
   B220-G275 &	P09 & 16.55  &  22.36$\;\;$ &  -5.14$\;\;$ &   22.95$\;\;$ &  0.06 &  -1.70 & 25.15  & 24.48 \\  
   B224-G279 &	P09 & 15.45  &  21.87$\;\;$ &  -7.34$\;\;$ &   23.07$\;\;$ &  0.07: &  -1.80: & 25.14:  & 24.45: \\  
   B225-G280 &  P09 & 14.15  &  16.48$\;\;$ & -12.19$\;\;$ &   20.50$\;\;$ &  0.05: &  -0.50: & 25.35:  & 24.55: \\  
   B233-G287 &	R05 & 15.76  &  35.45$\;\;$ &  -0.20$\;\;$ &   35.45$\;\;$ &  0.10 &  -1.53 & 25.25  & 24.43 \\  
   B240-G302 &	R05 & 15.21  &  11.02$\;\;$ & -29.81$\;\;$ &   31.78$\;\;$ &  0.14 &  -1.66 & 25.23  & 24.34 \\  
   B292-G010 &	P11 & 16.99  & -58.48$\;\;$ &  47.17$\;\;$ &   75.13$\;\;$ &  0.15 &  -1.90 & 25.39  & 24.50 \\  
   B293-G011 &	R05 & 16.29  & -61.86$\;\;$ &  43.64$\;\;$ &   75.70$\;\;$ &  0.12 &  -1.70 & 25.25  & 24.44 \\  
   B298-G021 &	M07 & 16.59  & -58.25$\;\;$ &  22.80$\;\;$ &   62.55$\;\;$ &  0.09 &  -1.80 & 25.17  & 24.49 \\  
   B311-G033 &  R05 & 15.44  & -57.57$\;\;$ &   0.99$\;\;$ &   57.57$\;\;$ &  0.25 &  -1.75 & 25.49  & 24.31 \\  
   B336-G067 &	P11 & 17.81  &  28.13$\;\;$ &  49.44$\;\;$ &   56.88$\;\;$ &  0.10 &  -1.90 & 25.28  & 24.50 \\  
   B337-G068 &	P11 & 16.73  &  30.99$\;\;$ &  51.44$\;\;$ &   60.06$\;\;$ &  0.06 &  -1.30 & 25.24  & 24.40 \\  
   B338-G076 &	R05 & 14.25  & -44.09$\;\;$ &  -9.05$\;\;$ &   45.01$\;\;$ &  0.04 &  -1.20 & 25.01  & 24.32 \\  
   B343-G105 &	R05 & 16.31  & -57.45$\;\;$ & -30.05$\;\;$ &   64.83$\;\;$ &  0.10 &  -1.50 & 25.41  & 24.61 \\  
   B350-G162 &	P11 & 16.74  & -42.29$\;\;$ & -29.21$\;\;$ &   51.40$\;\;$ &  0.11 &  -1.80 & 25.25  & 24.45 \\  
   B358-G219 &  R05 & 15.22  & -64.55$\;\;$ & -58.61$\;\;$ &   87.19$\;\;$ &  0.05 &  -1.91 & 25.16  & 24.62 \\  
   B366-G291 &	P09 & 15.99  &  51.62$\;\;$ &  11.50$\;\;$ &   52.88$\;\;$ &  0.09 &  -1.80 & 25.30  & 24.53 \\  
   B379-G312 &	R05 & 16.18  &  -3.67$\;\;$ & -49.65$\;\;$ &   49.79$\;\;$ &  0.13 &  -0.50 & 25.50  & 24.25 \\  
   B384-G319 &	R05 & 15.75  & -20.90$\;\;$ & -69.01$\;\;$ &   72.10$\;\;$ &  0.04 &  -0.50 & 25.33  & 24.46 \\  
   B386-G322 &	R05 & 15.55  &  61.67$\;\;$ &  -4.30$\;\;$ &   61.82$\;\;$ &  0.04 &  -1.10 & 25.16  & 24.46 \\  
   B405-G351 &	R05 & 15.19  &  63.69$\;\;$ & -48.84$\;\;$ &   80.26$\;\;$ &  0.08 &  -1.55 & 25.38  & 24.60 \\  
   B407-G352 &  P09 & 16.05  &  71.53$\;\;$ & -49.72$\;\;$ &   87.11$\;\;$ &  0.10 &  -0.40 & 25.41  & 24.35 \\  
   B468      &	R05 & 17.79  & -66.18$\;\;$ & -58.58$\;\;$ &   88.39$\;\;$ &  0.06 &  -0.70 & 25.41  & 24.45 \\  
   B514-MCGC4&  G06 & 15.76  &-242.32$\;\;$ & -15.11$\;\;$ &  242.79$\;\;$ &  0.09 &  -1.91 & 25.14  & 24.48 \\  
   B531      &	P11 & 19.63  & -59.10$\;\;$ &  47.17$\;\;$ &   75.62$\;\;$ &  0.14 &  -0.40 & 25.58  & 24.33 \\  
   B255D-D072&  P09 & 18.97  &  53.69$\;\;$ &  12.71$\;\;$ &   55.17$\;\;$ &  0.14 &  -0.70 & 25.50  & 24.38 \\  
   MCGC1-B520&  M07 & 16.05  &-182.12$\;\;$ &  91.29$\;\;$ &  203.72$\;\;$ &  0.12 &  -2.15 & 25.17  & 24.45 \\  
   MCGC2-H4  &  M07 & 16.98  & -90.47$\;\;$ & 115.36$\;\;$ &  146.61$\;\;$ &  0.10 &  -1.90 & 25.02  & 24.30 \\  
   MCGC3-H5  &	M07 & 16.31  & -67.02$\;\;$ & 122.53$\;\;$ &  139.67$\;\;$ &  0.10 &  -1.90 & 25.05  & 24.35 \\  
   MCGC5-H10 &	M07 & 16.09  &-315.17$\;\;$ &-141.46$\;\;$ &  345.46$\;\;$ &  0.11 &  -1.90 & 25.26  & 24.50 \\  
   MCGC7-H14 &	M07 & 18.27  &  25.99$\;\;$ &  75.37$\;\;$ &   79.73$\;\;$ &  0.06 &  -0.70 & 25.18  & 24.20 \\  
   MCGC8-H23 &	M07 & 16.72  &  11.26$\;\;$ &-162.31$\;\;$ &  162.70$\;\;$ &  0.09 &  -1.53 & 25.30  & 24.53 \\  
   MCGC9-H24 &	M07 & 17.78  & 161.35$\;\;$ & -55.63$\;\;$ &  170.67$\;\;$ &  0.16 &  -1.40 & 25.38  & 24.25 \\  
   MCGC10-H27 & M07 & 16.50  & -66.57$\;\;$ &-435.83$\;\;$ &  440.88$\;\;$ &  0.09 &  -1.90 & 25.15  & 24.50 \\  
   G001-MII   & R05 & 13.81  &-149.69$\;\;$ &  29.32$\;\;$ &  152.54$\;\;$ &  0.04 &  -0.90 & 25.23  & 24.56 \\  
   MCEC1-HEC5 & M06 & 17.60  &  -5.64$\;\;$ &  58.24$\;\;$ &   58.51$\;\;$ &  0.10 &  -1.91 & 25.17  & 24.48 \\  
   MCEC2-HEC7$^1$ & M06 & 17.10  & 128.39$\;\;$ &  97.87$\;\;$ &  161.44$\;\;$ &  0.13 &  -1.75 & 25.31  & 24.51 \\  
   MCEC3-HEC4$^1$ & M06 & 17.60  & -57.17$\;\;$ &  22.57$\;\;$ &   61.47$\;\;$ &  0.09 &  -1.91 & 25.08  & 24.42 \\  
   MCEC4-HEC12& M06 & 18.84  & -36.22$\;\;$ &-261.42$\;\;$ &  263.92$\;\;$ &  0.11 &  -1.78 & 25.14  & 24.33 \\  
\noalign{\smallskip}  							     
\hline		      							      
\end{tabular}	      							     
\end{center}	      
\tablefoot{ $^1$ In the RBCV4.0 the alternate names for MCEC2 and MCEC3 are erroneously 
switched, i.e.  HEC4 and HEC7 respectively.
} 
\end{table*}

\subsection{The colour-magnitude diagrams}

The CMDs of the 48 GCs considered in this paper were all obtained 
by our research group, using $HST$ data taken with the Faint Object Camera (FOC, one cluster 
only, B405), Wide Field Planetary Camera 2 (WFPC2)  and Advanced Camera for Surveys (ACS) 
from our own observing programs, or extracted from the $HST$ archive if 
observed by other programs.\footnote{The $HST$ photometric data were 
converted to the BVI magnitudes of the Johnson standard system according  
to FFP96 for the FOC F430W (B) and F480LP (V) filters, Holtzman et al. (1995)  
for the WFPC2 F555W (V) and F814W (I) filters, and Sirianni et al. (2005)  
for the ACS F606W (V) and F814W (I) filters.}
They  can be divided into two subsets: 
i) those that were observed and processed by us using the data reduction 
photometric package ROMAFOT (18 clusters from R05); 
ii) those that were processed by us using DOLPHOT (one external cluster from G06; 
11 clusters from  P09; 5 clusters from  P11, 4 extended clusters from Mackey 
et al. 2006, hereafter M06; 9 clusters in the external regions of M31 from 
Mackey et al. 2007, hereafter M07).   

The CMDs of the clusters can be considered as homogeneous within the errors 
even though they were processed with different packages and 
somewhat different procedures.
A special case is G011, the only cluster that has been
observed twice with the WFPC2 V and I bands. The first set of data, 
obtained in 1999 to study its CMD, was processed with ROMAFOT and 
the resulting CMD, presented by R05, is the one used in the present analysis. 
The second set of data, 
obtained in 2007 to study the variable stars, was processed with HSTPHOT  
and the resulting CMD will be presented by Contreras et al. (2012). 
These two CMDs are entirely equivalent despite the different photometric 
accuracy and scatter around the ridge lines, and support our previous 
claim that all CMDs processed by our group can be considered as 
homogeneous within the errors, irrespective of the packages and procedures 
applied in the data reduction and analysis. 

The CMDs of the 13 clusters observed and studied by M06 and M07 
were re-derived in the present study by  applying the same data reduction 
and analysis criteria and procedures as described e.g. by P09 to the $HST$ archive data. 
The resulting CMDs  are, as expected, very similar to those obtained by M06 
and M07, and hence our values for the metallicity, distance modulus and extinction 
are comparable to the M06 and M07 results.  The re-reduction step was nevertheless 
necessary to ensure the complete homogeneity of our database. 

The different procedures in the original determination of the respective 
V(HB) level are superseded by the homogeneous procedure applied to all clusters 
in this study (see Sect. \ref{ss:vhb}). 

The CMDs were all decontaminated from the field contribution, except 
seven (MCGC1, MCGC2, MCGC3, B514, MCGC5, MCGC9, MCGC10) for which the field 
contamination is negligible,  and three (B158, B220 and B224) for which the 
decontamination procedure is statistically unreliable 
because of the strong difference in completeness between the very crowded 
cluster and the field.  In several clusters, blended stars in the innermost 
regions were also eliminated. The cleaned CMDs are  shown in 
Fig. \ref{f:cmda}. 
The original observed  CMDs have been published in the quoted papers.






\begin{figure*}
 \centering\includegraphics[width=17.6cm]{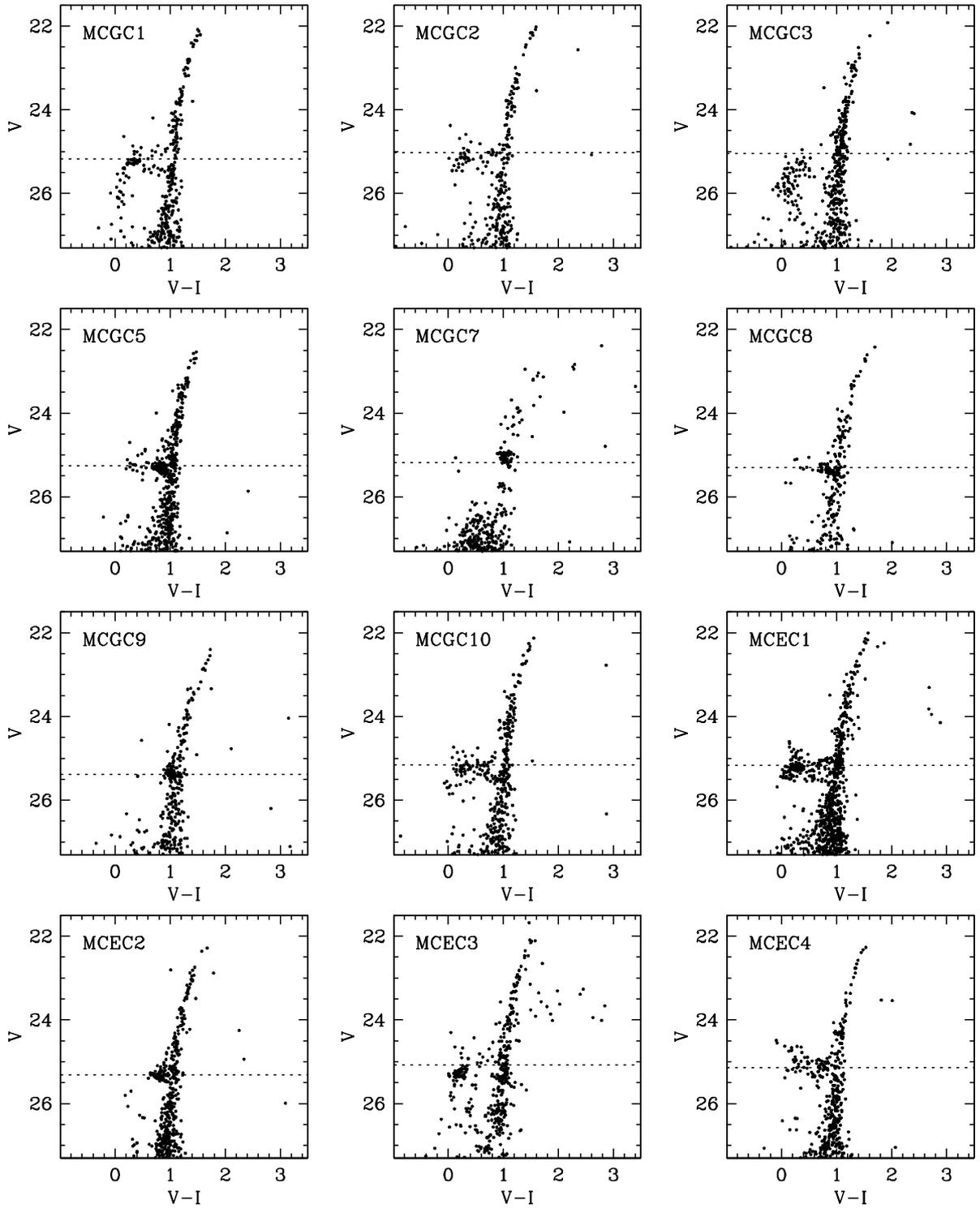}
 \caption{Colour-magnitude diagrams of the target clusters. The dotted lines represent 
the level of the measured V(HB).}
\label{f:cmda}
\end{figure*}

\addtocounter{figure}{-1}

\begin{figure*}
 \centering\includegraphics[width=17.6cm]{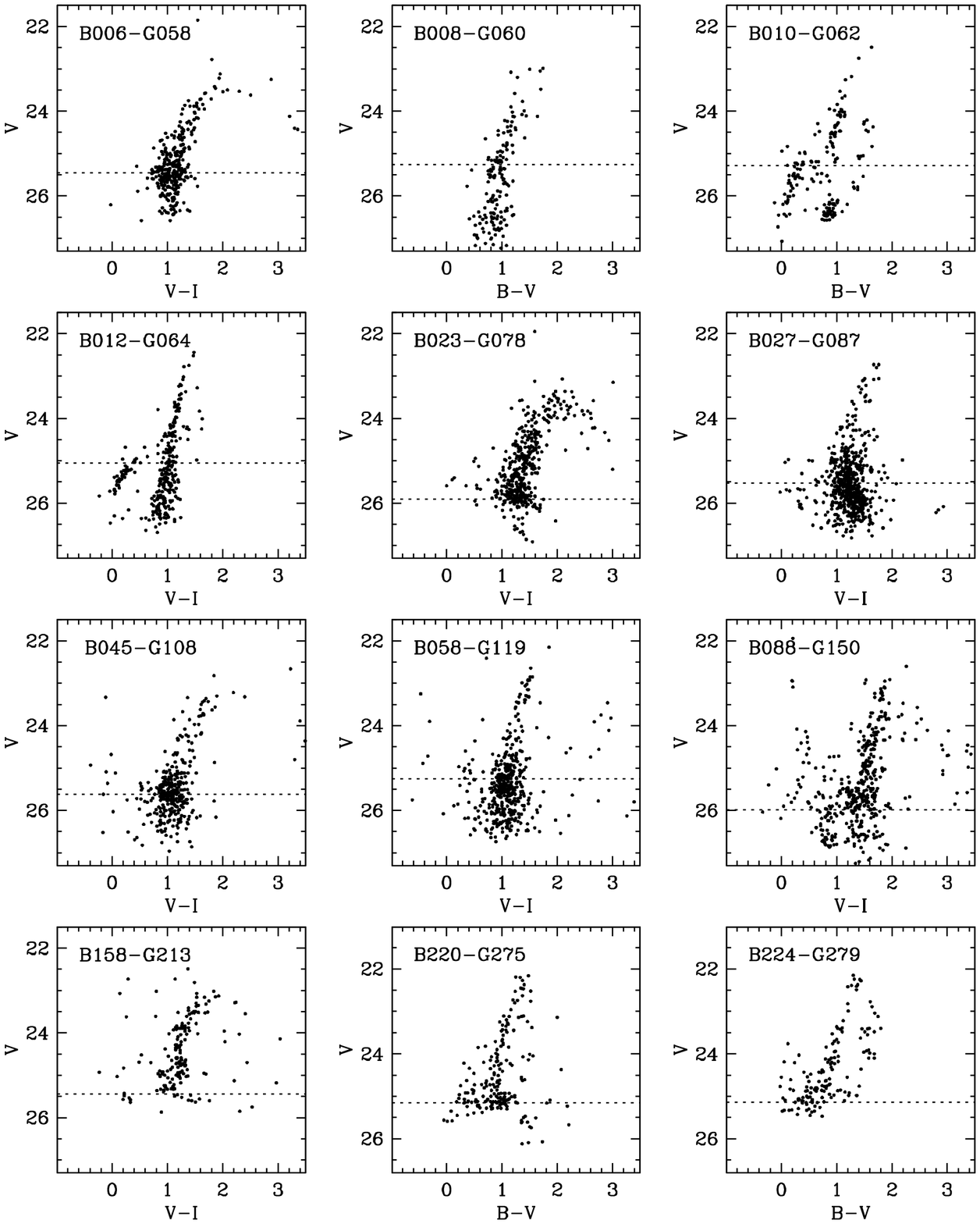}
\caption{-- continued}
\label{f:cmdb}
\end{figure*}

\addtocounter{figure}{-1}

\begin{figure*}
 \centering\includegraphics[width=17.6cm]{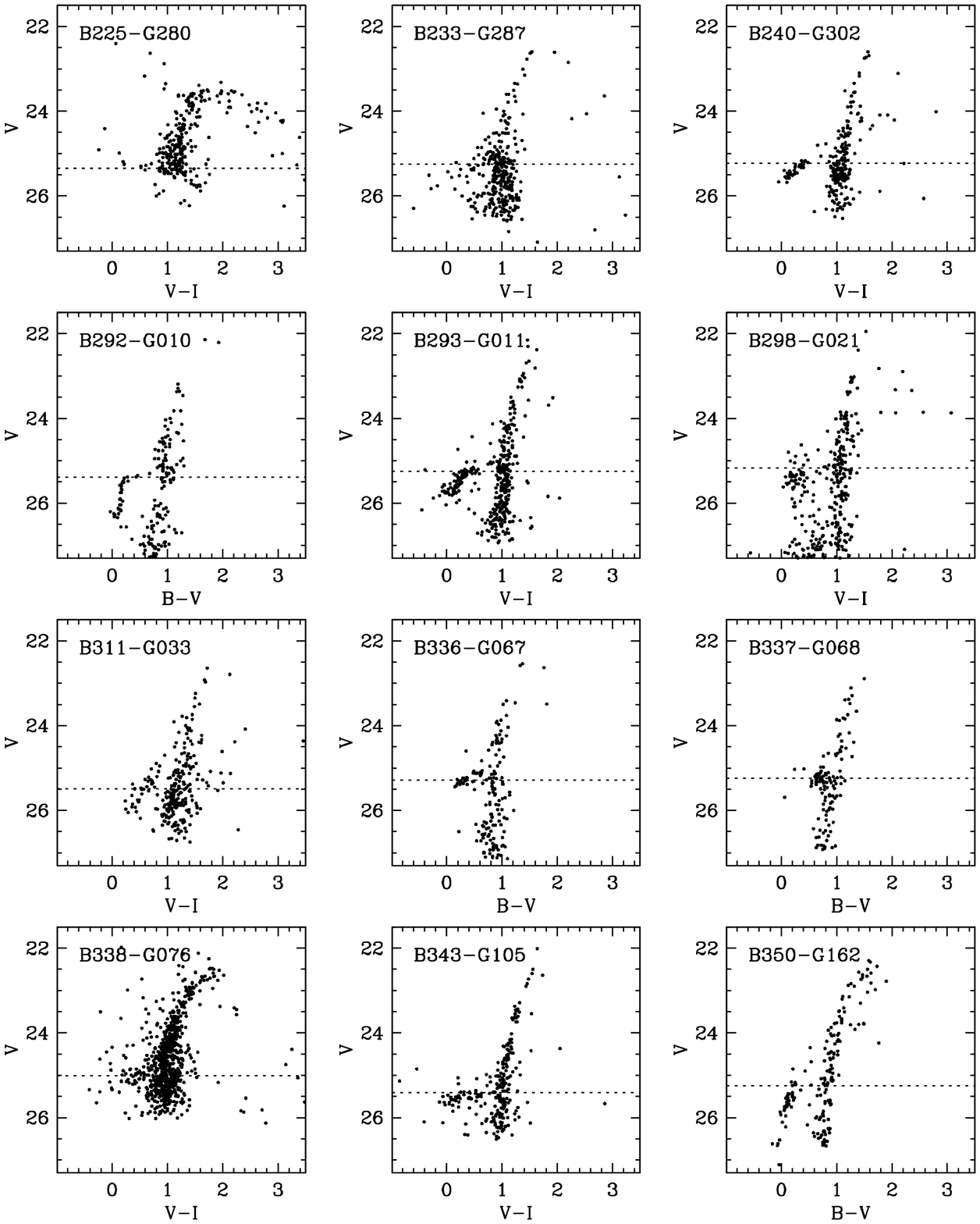}
\caption{-- continued}
\label{f:cmdc}
\end{figure*}

\addtocounter{figure}{-1}

\begin{figure*}
 \centering\includegraphics[width=17.6cm]{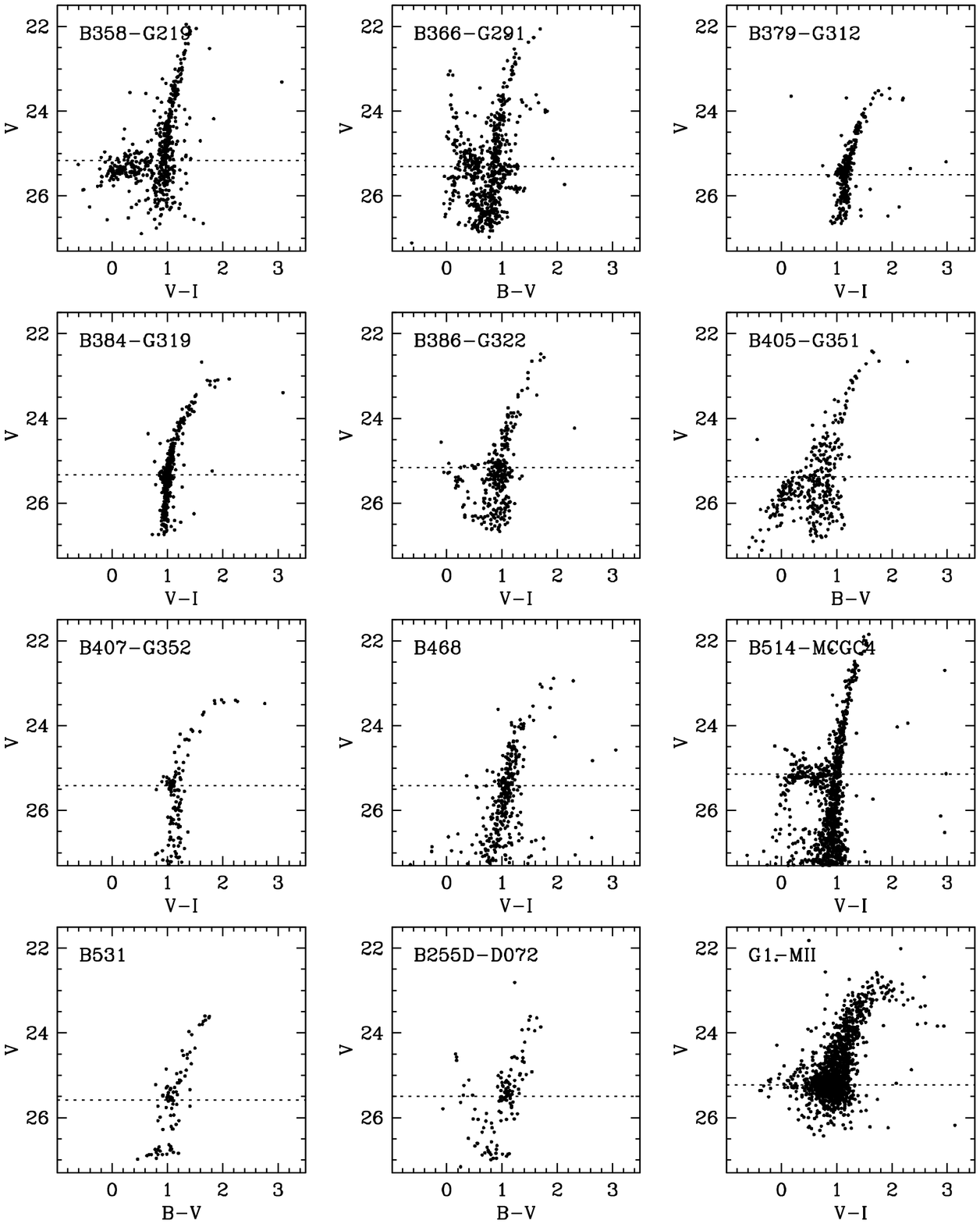}
\caption{-- continued.}
\label{f:cmdd}
\end{figure*}

\section{Reddening, metallicity and distance from the CMD}\label{s:redmetlum} 

\subsection{The method}\label{ss:method}

As described in detail e.g. in R05, M07, P09 and P11, estimates of reddening, 
metallicity 
and distance are obtained simultaneously by comparing the observed CMDs of the 
M31 GCs with the CMD ridge lines of a set of reference Galactic GCs that are 
selected to sample a wide range of metallicity. 
We briefly recall here the basic 
steps of this procedure, and refer the reader to P09 for a more detailed 
description and discussion.

\begin{table}[t]
\caption{Reference grid of template Galactic globular clusters.  
\label{t:grid}}
\begin{center}
\begin{tabular}[t]{lcccc}
\hline\noalign{\smallskip}
ID   & [Fe/H]$_{ZW}$  & E(B--V) & $\mu _V$ & Phot.  \\
     &  dex           &         &   mag    &        \\
\noalign{\smallskip}
\hline\noalign{\smallskip}
NGC6341 (M92) & -2.24 &  0.031  &   14.84  &  VI    \\                                     
NGC7078 (M15) & -2.15 &  0.084  &   15.51  &  BV,VI \\                                     
NGC4147       & -1.80 &  0.018  &   16.48  &  BV    \\                                        
NGC5272 (M3)  & -1.66 &  0.018  &   15.11  &  VI    \\                                       
NGC6205 (M13) & -1.65 &  0.019  &   14.50  &  BV    \\                                          
NGC5904 (M5)  & -1.40 &  0.034  &   14.43  &  BV,VI \\                                         
NGC6723       & -1.12 &  0.044  &   14.73  &  BV    \\
NGC104 (47Tuc)& -0.71 &  0.023  &   13.33  &  BV,VI \\
NGC5927       & -0.30 &  0.399  &   15.81  &  BV,VI \\
\noalign{\smallskip}  							     
\hline		      							      
\end{tabular}	      							     
\end{center}	      
\tablefoot{Metallicities are from Zinn (1985);  VI photometry is from Rosenberg
et al. (2000a,b); BV photometry is from Piotto et al. (2002). Reddening and distance 
moduli are from Dotter et al. (2010). 
}
\end{table}	

\begin{itemize}      
		      
\item
The reference Galactic GCs (GGC) used in the present study are listed in 
Table \ref{t:grid}.    
They sample a metallicity range of [Fe/H]= --0.30 to --2.24 dex  
where the values 
of metallicity are taken from Zinn (1985) in the Zinn and West (1984, 
hereafter ZW84) 
metallicity scale for homogeneity with the previous studies. 
The CMD ridge lines of the template clusters are transferred to the 
M$_V$ vs. intrinsic colour plane by using the reddening values and the distance 
moduli in Table \ref{t:grid}, which were taken from Dotter et al. (2010). 
The reasons for choosing this database rather than Harris' (1996) database 
of MW GCs properties is discussed in detail in Sect. \ref{ss:refGC}. 


\item  
We searched for the set of parameters (distance, reddening and metallicity) for 
each M31 GC leading to the best match between its observed RGB and HB and the 
corresponding ridge lines of the reference clusters, according to the relations: 
A$_V$ = 3.1E(B -- V), A$_I$ = 1.94E(B -- V) and E(V -- I) = 1.375E(B -- V) 
(Schlegel et al. 1998).
Colour and magnitude shifts are applied iteratively to the observed RGB and HB 
until a satisfactory match with a template is found.

\item From these shifts we simultaneously estimated reddening and distance, while 
the metallicity was estimated by interpolation between the two RGB template lines 
that bracket the observed RGB. The best fit with the HB ridge lines leads also 
to an estimate of V(HB), as described in more detail in Sect. \ref{ss:vhb}.

\end{itemize} 

The best fit (by eye) of the observed and template CMDs, which implies 
a shift in magnitude (due to distance and absorption) and a shift in colour 
(due to reddening and metallicity) is not quite trivial and requires a 
careful evaluation and a number of iterations.  
However,  in most cases, where the various branches of the CMD are well 
populated and defined, the pairs of reddening and metallicity values can be 
derived with a good degree of confidence and reliability. 
The values  obtained with this procedure are listed in Table \ref{t:targets}. 
Typical errors are about $\pm$0.04 mag for reddening,  $\pm$0.25 dex for 
metallicity and $\pm$0.10 mag for V(HB),  except for the clusters B023, 
B027, B158, B224 and B225 where the HBs are not well defined and hence these 
estimates, in particular  V(HB), are more uncertain. 
These clusters were not used in the analysis described in Sect. \ref{s:hbmet}, 
but are shown for comparison along with the results obtained for the remaining 
43 clusters. 
 
Estimates of reddening and metallicity were derived in previous studies 
with spectroscopic and photometric techniques, and we compare them with 
our results in the following subsections.

\subsection{Reddening}\label{ss:redd}  

A detailed description of previous reddening estimates for individual GCs in M31 
is given in R05. 
More recent studies are those by i) Fan et al. (2008), based on correlations 
between optical and infrared colours and metallicity, combined with the use of  
various reddening-free parameters (as in Barmby et al. 2000); 
ii) Montalto et al (2009), based on a multiwavelength (far UV to IR) photometric 
study of dust properties; and iii) Caldwell et al. (2011), based on 5{\AA} resolution  
spectra of target clusters compared to flux-calibrated spectra of reference clusters 
with similar metallicity that were dereddened using the Barmby et al. (2000) values.  

\begin{figure}
\centering
\includegraphics[width=9.0cm]{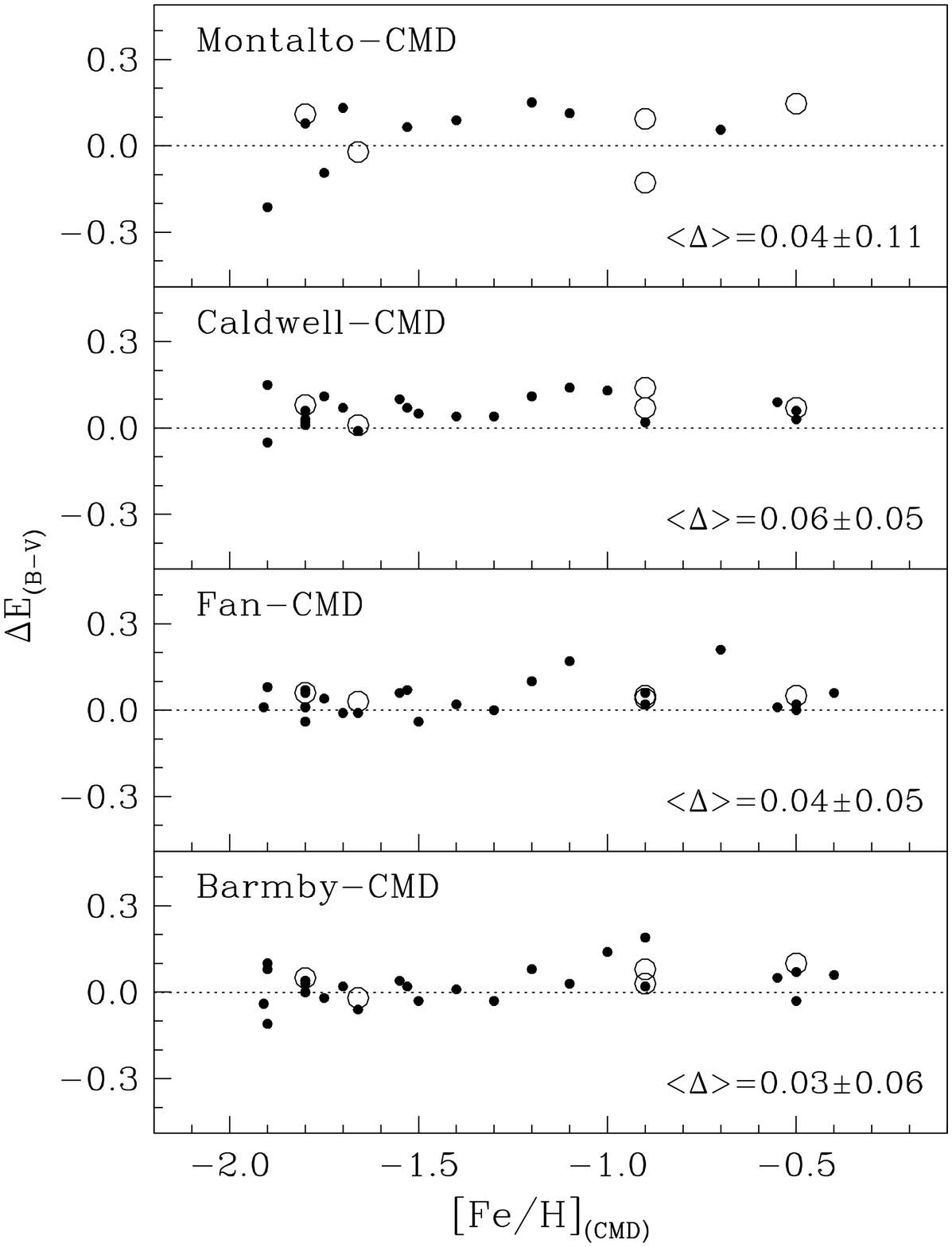}
\caption{Comparison of our reddening estimates from CMD fitting with 
the most recent reddening estimates (Fan et al. 2008; 
Montalto et al. 2009; Caldwell et al. 2011) as well as with those of Barmby et al. (2000), 
who provided the calibration frame for the Fan et al. and Caldwell et al. results.
The open circles indicate the clusters excluded from the analysis in Sect. \ref{s:hbmet}, 
for completeness (see Sect. \ref{ss:method}).
}
\label{f:ebmv}
\end{figure}

The comparison of our 
results with these studies as a function of metallicity is shown in Fig. \ref{f:ebmv}. 
The values estimated by  Fan et al. (2008) and Caldwell et al. 
(2011) do not show any trend with respect to ours, and are offset by 
0.04$\pm$0.05 and 0.06$\pm$0.05 mag, respectively, reproducing the general pattern 
of Barmby et al. (2000) results, which were used as calibrating reference frame 
in both these studies. 
The  Montalto et al. (2009) results, which are based on an entirely independent 
study of dust properties, show a larger scatter and overestimate the reddening with 
respect to ours by 0.04$\pm$0.11 mag. This might suggest that the 
dust properties considered by Montalto et al. (2009) are not well represented by the 
photometric and spectroscopic properties of the cluster integrated light. 

The impact of systematics in the reddening determination on the results of the 
present analysis is discussed in Sect. \ref{ss:ass-redd}.

\subsection{Metallicity}\label{ss:met}

\begin{figure}
\centering
\includegraphics[width=9.0cm]{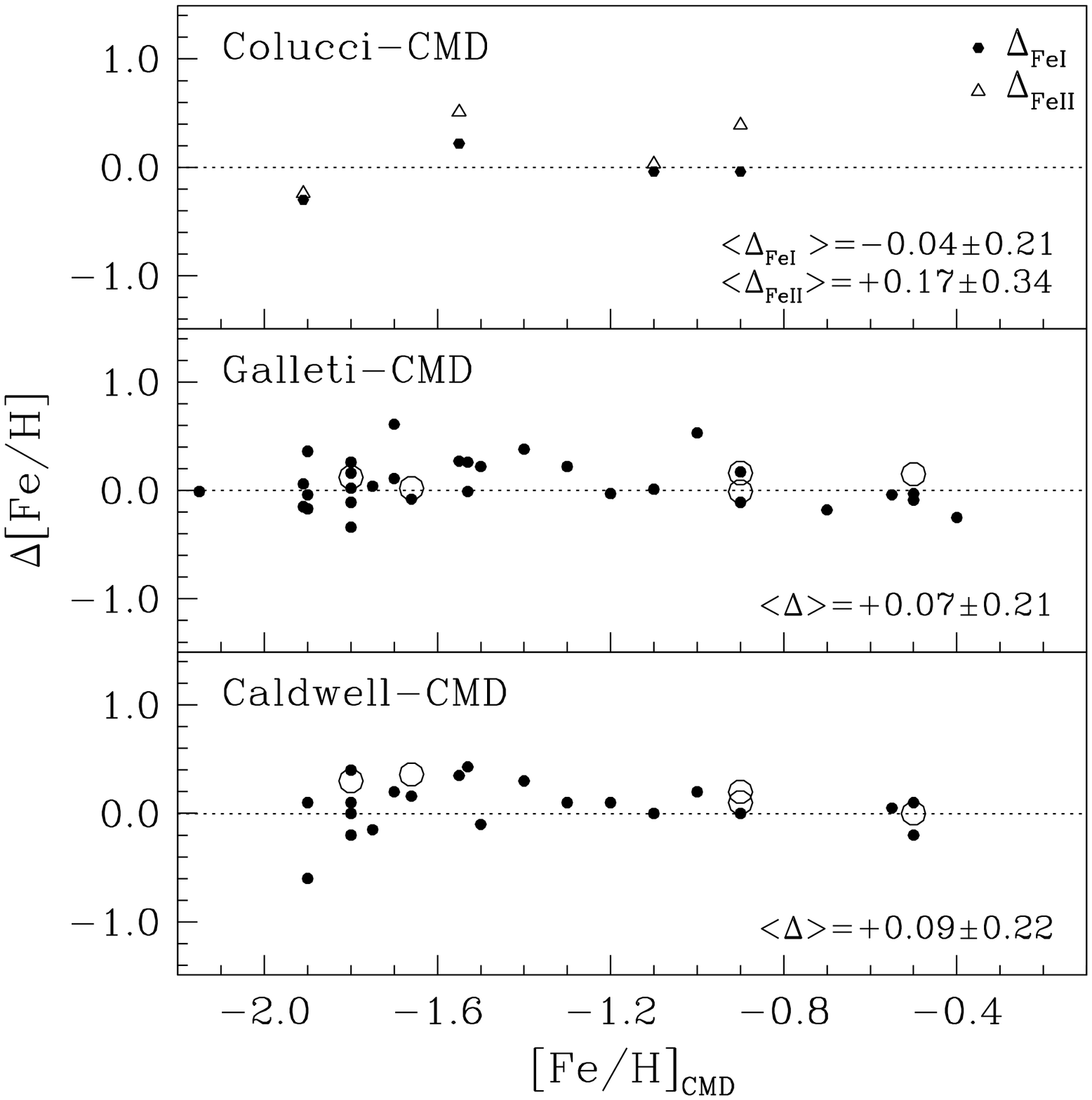}
\caption{Comparison of our metallicity estimates from CMD fitting with 
the most recent metallicity estimates (Galleti et al. 2009; 
Colucci et al. 2009; Caldwell et al. 2011). 
The open circles indicate the clusters excluded from the analysis in Sect. \ref{s:hbmet}, 
for completeness (see Sect. \ref{ss:method}).
}
\label{f:dmet}
\end{figure}

Previous estimates of metallicity have been described and discussed in detail 
by R05. 
Additional metallicity values have been obtained more recently, by Galleti 
et al. (2009), Colucci et al. (2009),  
and Caldwell et al. (2011). The Galleti et al. (2009) 
and  Caldwell et al. (2011) estimates have been obtained from  spectro-photometric 
Lick indices calibrated on well-studied Galactic GCs.
We show in Figure \ref{f:dmet} the comparison of these values with our 
estimates from CMD fitting, which appear to be systematically underestimated 
by $\sim$ 0.09-0.12 dex, well within the r.m.s. error of $\sim$ 0.25 dex of 
these estimates.

On the other hand, Colucci et al. (2009) obtained high-resolution spectra of 
five M31 GCs and derived metallicities from Fe$_{I}$ and Fe$_{II}$ lines, 
as well as other elements. Four of these five clusters are included in 
our database, and we show in Fig. \ref{f:dmet} that the results based 
on  Fe$_{I}$ are virtually identical to ours 
($\Delta$[Fe/H]=0.02$\pm$0.22 dex). Moreover, there is a 
systematic difference of $\sim$ 0.2 dex between the estimates based 
on Fe$_{I}$ and Fe$_{II}$ lines. This is an interesting piece of information, 
because the metallicites based on Fe$_{I}$ lines are considered to be more robust because 
of the larger number of lines and the lower dependence on gravity. 
It also stresses the importance of using homogenous values 
for the parameters involved in global comparisons, in particular for the 
metallicity scale. 

The impact of systematics in the metallicity determination on the results of the 
present analysis is discussed in Sect. \ref{ss:ass-met}.

\subsection{The HB magnitude level V(HB)}\label{ss:vhb}

The HB is particularly important in a GC CMD, because it gives information 
on two fundamental subjects: i) the characteristics and evolution of the 
stellar component(s), via the HB morphology;  and ii) the cluster distance, 
via the HB luminosity. The matter of the HB morphology and its implications 
will be treated in a forthcoming paper (Perina et al. 2012, in preparation), 
here we only deal with the HB luminosity level as a distance indicator.

In the present study as in P09, we estimated the value of 
V(HB) using the magnitude level of the best-fitting template HB ridge 
line at (V--I)$_0$=0.5 or (B--V)$_0$=0.3, corresponding to the middle of the 
instability strip, for the intermediate and metal-poor clusters. 
The reasons for applying this procedure are that in several clusters 
the HB population is quite low and a running box procedure (which we used e.g. 
in R05) is less reliable, and that we aim at deriving all quantities 
(reddening, 
metallicity, distance) from the application of the same self-consistent CMD 
best-fitting method. 

In the metal-rich clusters the HB is only populated on the red clump (RC), 
which is easily identifiable on the RGB luminosity function.  
According to stellar evolution theory (e.g. Lee et al. 1994), the RC is 
slightly brighter than the average HB luminosity within the instability strip, 
if RR Lyrae variables were present in these metal-rich clusters.   
For consistency with the intermediate and metal-poor clusters discussed above, 
we need to correct the V(RC) mag  to transfer it to the level of the 
corresponding theoretical instability strip. The size of this correction ranges  
from $\sim$0.05 to 0.12 mag according to different Zero Age Horizontal Branch (ZAHB) 
models (see FFP96 for 
more details), and we adopted a correction of +0.08 mag for the clusters 
with [Fe/H]$>$--1.0 dex, for consistency 
with most previous studies (Sarajedini et al. 1995; Ajhar et al. 1996; Catelan 
\& de Freitas Pacheco 1996) as well as with our own studies (FFP96, R05, P09). 
We note, however, that in some cases (e.g. Harris 1996, 2010 edition of his GGC 
online Catalogue)\footnote{http://www.physics.mcmaster.ca/Globular.html}  
this correction is not applied, and what appears as V(HB) in the metal-rich 
clusters is actually the mean V(RC).  

The values of V(HB) estimated and adopted for the present analysis 
are listed in Table \ref{t:targets}.  
 
In conclusion, our estimates of reddening and metallicity obtained with the 
CMD-fitting method described above generally agree well 
with most previous estimates based on different methods. 
Since the aim of the present work is to ensure the homogeneity of our 
data set, we  adopt in the following analysis the self-consistent 
set of values of reddening, metallicity and distance obtained from our 
CMD fitting method, which are listed in Table \ref{t:targets}.

\section{The HB luminosity-metallicity relation}\label{s:hbmet}

\begin{figure}
\centering
\includegraphics[width=9.0cm]{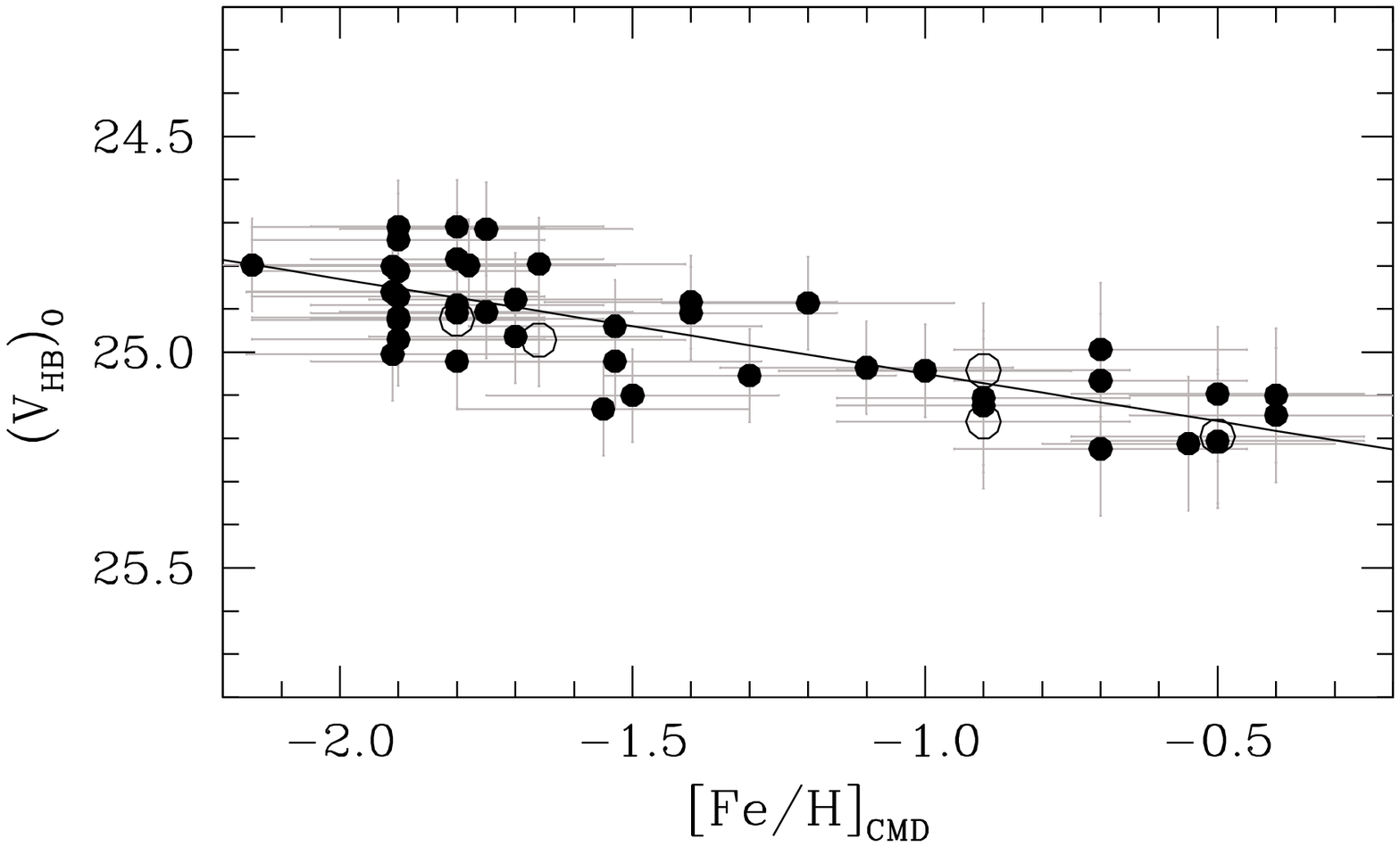}
\caption{V$_0(HB)$ estimates as a function of [Fe/H] (in the ZW84 metallicity 
scale) for the present sample of 48 M31 GCs. The line represents the best linear fit  
as expressed in Eq. (1). The open symbols show the GCs that were excluded from the 
analysis in Sect. \ref{s:hbmet} (see Sect. \ref{ss:method}). 
}
\label{f:vhbfe}
\end{figure}

We show in Fig. \ref{f:vhbfe} the distribution of the V(HB)$_0$  
values as a function of [Fe/H] for our 48 target clusters using the 
data in Table \ref{t:targets}. 
Because there are errors in both V(HB)$_0$ and [Fe/H], we have applied 
an orthogonal least-squares program to determine the linear 
regression, and jackknife resampling simulations for the error 
analysis of the relation  (Feigelson \& Babu 1992). The five 
GCs mentioned in Sect.  \ref{ss:method} were not used in this 
fitting procedure, and are shown as open symbols for completeness and 
comparison. We find: 
\begin{equation}
V_0(HB) = 0.22 (\pm 0.02)[Fe/H] + 25.27 (\pm 0.04).   
\end{equation}
This relation does not change significantly when  the five GCs that 
were left out are included, therefore our procedure and results appear to be  
sufficiently reliable and accurate, within the errors, even in 
those cases where the CMD is less well defined.  
In this relation the individual cluster distances are not taken into account, 
therefore in addition to the errors on photometry and metallicity determination, 
the scatter of the distribution is partly  due to line-of-sigth depth effects,  
because some of the most discrepant clusters 
may be located significantly far away in the background or foreground with respect 
to the main M31 cluster population (see e.g. M07). 
The reddening also contributes to the scatter, especially when the CMD is not 
defined sufficiently well to break the reddening-metallicity degeneracy. 

\begin{figure}
\centering
\includegraphics[width=9.0cm]{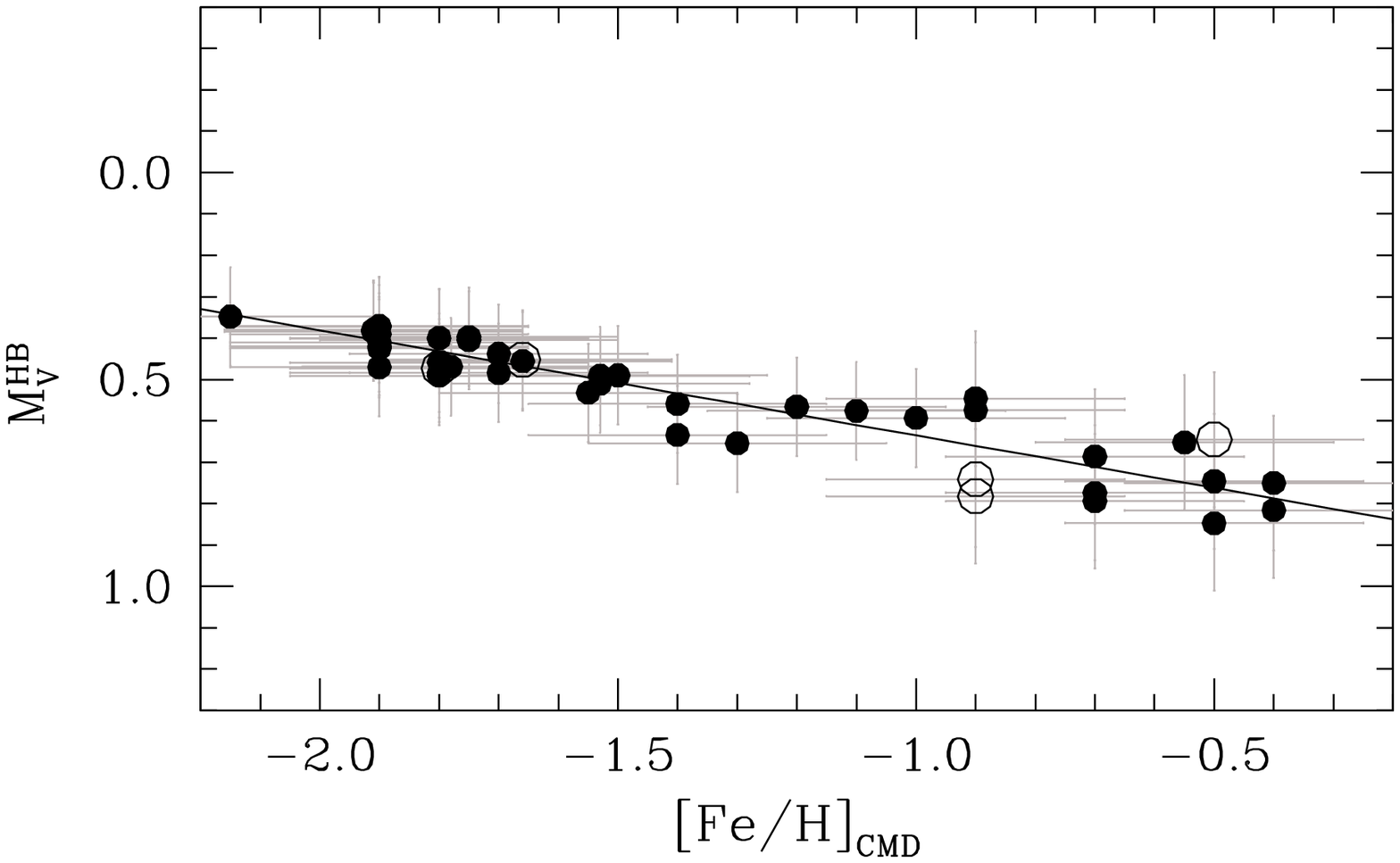}
\caption{M$_V$(HB) estimates as a function of [Fe/H] (in the ZW84 metallicity 
scale) for the present sample of 48 M31 GCs. The line represents the best linear fit  
as expressed in Eq. (2).  The open symbols show the GCs that were excluded from the 
analysis in  Sect. \ref{s:hbmet} (see Sect. \ref{ss:method}).  
}
\label{f:Mvhbfe}
\end{figure}

By considering the individual cluster distance moduli derived from the CMD fitting
procedure that are listed in Table \ref{t:targets}, and using the same statistical 
procedure as for eq. (1), we obtain the relation: 
\begin{equation}
M_V(HB) = 0.25 (\pm 0.02)[Fe/H] + 0.89 (\pm 0.03),   
\end{equation}
which 
is shown in Fig. \ref{f:Mvhbfe}.

Eq. (2) is obviously better defined than eq. (1) because the scatter due to 
line-of-sight depth effects has been corrected for. This shows and confirms 
that the method we used for our analysis is able to estimate the individual 
cluster distances sufficiently well to significantly reduce the scatter in the distribution 
of a collective sample property. The different slopes of eq.s (1) and (2) are 
well compatible within the errors. 
It is worth noting here that without  the correction of 0.08 mag applied 
to the observed V(HB) in metal-rich GCs (see Sect. \ref{ss:vhb}) the slope 
of eq.s (1) and (2) would be flatter ($\sim$ 0.15-0.19, respectively).   
  
By normalising eq.s (1) and (2) near the middle of the metallicity range, 
i.e. at [Fe/H]=--1.5 for convenience, where $V_0(HB)$=24.94$\pm$0.05 mag from
eq.(1) and $M_V$(HB)=0.52$\pm$0.04 mag  from eq.(2),  
we derive a distance modulus of 24.42$\pm$0.06 mag for M31. 
This result can be regarded as an independent distance determination to M31 based 
on the distance scale set by the Milky Way GCs. It agrees excellently  
with the average value of the distances to M31 derived during the past two decades 
using Cepheids, 
carbon-rich stars, 
TRGB and RC stars, 
and eclipsing binaries 
(cf. Vilardell et al. 2006; 2010, and references therein), and in particular with the Riess et al. (2012)
result based on 68 classical Cepheids. 

We therefore propose eq. (2) as a reliable result of our analysis,  
which leads to a distance modulus for the LMC of 18.54$\pm$0.07  mag based on the
Clementini et al. (2003) RR Lyrae data ($<V_0(RR)>$ = 19.064 $\pm$ 0.064 mag 
at [Fe/H] = --1.5). 
This result agrees very well with the most recent calibration of 
the $M_V$(RR)-[Fe/H] relation (Benedict et al. 2011), which is based on $HST$ 
trigonometric parallaxes for a few classical Cepheids 
and RR Lyrae variables in the MW as well as on other distance estimates, and
yields a distance modulus for the LMC of 18.55$\pm$0.05 mag. 

A review by Clementini (2009) of the distance determinations to the LMC obtained during 
the past decade using pulsating variable stars lists values of (m-M)$_0$ 
ranging from 18.39 to 18.58 mag with a mean value of 18.52 $\pm$ 0.01 mag 
(internal rms error only). Quasi-geometric distances from detached eclipsing binary 
systems (Fitzpatrick et al. 2003; Pietrzynski et al. 2009; Bonanos et al. 2011) 
lead to a mean distance modulus of 18.49 $\pm$ 0.04 mag. 
Several of these individual studies quote internal errors of less than 0.05 mag, so we 
are presently in a situation where systematic (calibration) errors dominate. 

A more accurate and robust determination of the distance to M31 (as well as to the LMC)   
is expected to be possible in the near future from ongoing developments, e.g.  
the study of Cepheids in the IR to minimise the effects of metallicity and reddening 
(Freedman et al. 2011), and direct water maser observations (Darling 2011). 
Significant improvement is also expected from all other distance determination methods 
thanks to better observations and calibrations, and eventually from Gaia parallaxes 
(for the LMC) and proper motions of point-like sources brighter than V$\sim$20 
combined with a galaxy rotation model (for M31).

\section{The impact of systematics on the $M_V-[Fe/H]$ relation}\label{s:syst}

\subsection{The $M_V-[Fe/H]$ relation of the reference GGCs}\label{ss:refGC}

To report the observed CMDs of the MW reference GCs to the absolute luminosity 
plane $M_V$, we used in previous studies the distance moduli from the Harris 
(1996) 2003 edition of his GGC online Catalogue, which were obtained by assuming 
the GGCs HB luminosity-metallicity relation  M$_V$(HB)=0.16[Fe/H]+0.84 
calibrated on several different distance determination methods.  
In the present study we instead used the Dotter et al. (2010) distances, which were    
obtained from the best fit of the GGC main sequences with theoretical isochrones. 
We preferred to use this set of data because of the better accuracy of 
the observed CMDs (from $HST$/ACS photometry), the homogeneity of the analysis 
and the independence of the distance determination from the HB luminosity.   
The HB luminosity-metallicity relation found by Dotter et al. (2010)\footnote{See 
their eq.(1) and its transformation to $M_V(HB)$=0.235([Fe/H]+1.6)+0.53} 
is {\em derived} as a result of an independent analysis, and provides a consistency 
check to their procedure.   
The fact that the Dotter et al. relation is quite compatible with ours within 
the errors confirms that 
the M31 and MW GCs are indeed of a similar nature and share similar evolutionary 
properties, as we assumed at the beginning of our work.  
{\em The only way to break the dependence on this assumption can 
be provided by individual M31 GC distance determinations based 
on geometric or trigonometric methods, whenever they will be available}.

\subsection{The assumed reddening}\label{ss:ass-redd}

The reddening can play an important role by affecting the value of $V_0(HB)$, and hence 
increasing the scatter and  mimicking a distance effect.  
 
The comparison with other reddening estimates, presented and discussed in Sect. 
\ref{ss:redd}, 
shows that our E(B--V) estimates are on average about 
0.03, 0.04 and 0.06 mag smaller than those by Barmby et al (2000), Fan et al. (2008) 
and Caldwell et al. (2011), respectively. 
The differences are within the combined errors of the quoted  estimates with ours, 
and hence are hardly significant, but for the sake of completeness we can estimate 
how an offset of -0.05 mag in our reddening values would affect our results: the 
correction for such an offset would make our colours bluer by the same amount  
and our  $V_0(HB)$ values brighter by $\sim$0.15 mag, thus shifting the overall 
metallicity distribution towards lower metallicities. 
This would lead to a worse (and in most cases very poor or impossible) match between 
the observed and the template RGBs because of the dependence of the RGB {\em shape} 
on metallicity.

\subsection{The assumed metallicity}\label{ss:ass-met}

Using a different {\em metallicity scale}, for example the scale derived 
by Carretta et al. (2009), leads to the relation 
\begin{equation}
M_V = 0.24 (\pm 0.02)[Fe/H] + 0.87 (\pm 0.02),
\end{equation}
which is basically identical to eq. (2) within the errors. 

With respect to other {\em metallicity estimates} of the target clusters (see 
Sect. \ref{ss:met} and Fig. \ref{f:dmet}), 
our results agrees very well with high-resolution spectroscopy  (Colucci 
et al. 2009), but seem to underestimate the metallicity by about 0.1 dex with 
respect to spectro-photometric studies (Galleti et al. 2009; Caldwell et al. 2011). 
Again, this discrepancy is hardly significant, because it is  well within the 1$\sigma$ 
error of $\pm$ 0.25 dex. 
In addition, a metallicity underestimate is at odds with the systematics 
on reddening (if significant) discussed in the previous section, which would rather 
require the metallicity to be overestimated. 

Therefore, the method of CMD best fitting, which simultaneously constrains reddening and 
metallicity, should ensure that systematic biases in our estimates, if any, are minimised 
within the quoted errors for these determinations.

\subsection{Comparison with other $M_V(HB)-[Fe/H]$ relationships}\label{ss:comp}

\subsubsection{From empirical methods}\label{sss:compe}

Various empirical methods have been applied during the past decades to estimate 
the absolute magnitude of the HB (or of the RR Lyrae) stars and its dependence on 
metallicity. The most widely used methods include the Baade-Wesselink (B-W) analysis 
of field RR Lyraes, statistical and trigonometric parallaxes of RR Lyrae and BHB 
field stars, GC distance determination via main-sequence fitting and 
hence luminosities for RR Lyrae and BHB members.    
We refer the interested reader to Cacciari \& Clementini (2003) for a review.

From the observational point of view, the mean magnitude level of the HB in any 
given cluster, as defined in Sect. \ref{ss:vhb}, is equivalent to  
the mean magnitude of the RR Lyrae stars in that cluster, i.e. V(HB)=V(RR). 

The dependence of this luminosity on metallicity ranged from a slope 
$\Delta M_V(RR) / \Delta [Fe/H] \sim$ 0.27-0.37 mag dex$^{-1}$ (Sandage 1993; 
McNamara 1997; Feast 1997; Reid 1997) to 0.13-0.23 mag dex$^{-1}$ 
(Fernley et al. 1998 and references therein; Chaboyer 1999; Gratton et al. 2004; R05). 
A fairly accurate and widely  quoted result is the slope of 0.214 $\pm$ 0.047  
from the photometric and spectroscopic study of about 100 RR Lyrae stars in the 
bar of the LMC by Clementini et al. (2003), who found that  a unique linear 
relation over the entire considered metallicity range was adequate to fit the data. 
The slope of our eq. (3)  agrees excellently with this estimate.   

The zero-point of this relation was determined according to the various methods
applied, and varied significantly ($\sim$0.2 mag) from one study to another, 
as shown by its impact on the LMC distance estimates (cf. Sect. \ref{s:hbmet}).  
In this respect, 
one also has to consider the intrinsic dispersion in M$_V$ due to evolutionary
effects (see next section \ref{sss:compt}). 
The width of the HB as a function of metallicity was estimated  
empirically from the inspection of 14 GGCs by Sandage (1993), who  
expressed it as  $\Delta V(ZAHB-HB)=0.05[Fe/H]+0.16$. 
This contributes to the error associated to the mean estimated values of 
V(HB) or V(RR), especially when relatively few stars are considered.

\subsubsection{From evolution and pulsation models}\label{sss:compt}

Evolutionary models of Zero-Age Horizontal Branch (ZAHB) stars as well as 
pulsation models of RR Lyrae variables can provide an estimate of $M_V(ZAHB)$ or 
$M_V(RR)$, respectively,  as a function of metallicity (Caputo 2011). 
In this case, however,  $M_V(ZAHB)$ and $M_V(RR)$ are not equivalent, because 
the ZAHB represents the lower envelope of the HB locus, where the stars  
spend less than $\sim$ 10\% of their total HB lifetime, the remaining time being 
spent off the ZAHB at 0.1-0.2 mag brighter luminosities. 
Therefore, the observed mean HB (including the RR Lyrae stars) is represented 
by the main body ($\sim$ 90\%) of the stellar population, which is in a more 
advanced stage of evolution, and hence brighter than the ZAHB. 
This is usually taken into account by correcting the theoretical $M_V(ZAHB)$
by a constant offset ($\sim$0.10 mag) or by a linear function of metallicity,
such as that derived empirically by Sandage (1993), which we quoted in  
Sect. \ref{sss:compe}.

Some theoretical studies of stellar evolution and pulsation have derived  
linear relationships between $M_V(ZAHB)$ or $M_V(HB)$ and metallicity, with 
slopes that  in general agree well with the empirical results described 
above (Caloi et al. 1997; Demarque et al. 2000; Marconi \& Clementini 2005).  
The zero-point of these theoretical relations can differ by as much as 0.15 mag 
owing to the different assumptions that affect the absolute calibration.  

These studies did not find  any clear evidence for a change in slope at 
[Fe/H] = --1.5. 
However, other studies have suggested that the $M_V(HB)-[Fe/H]$ 
relation could be better approximated by a non-linear function.   
The non-linearity can be expressed as two linear relations  
changing slope at the breaking point [Fe/H] $\sim$ --1.5, as proposed 
by McNamara (1999) based on RR Lyrae empirical data, and by Caputo et al. (2000) 
based on stellar pulsation models. Both studies found that the metal-poor part 
of this relation was less steep (nearly flat according to McNamara) than the 
metal-rich  part. Alternatively, a quadratic relation  between $M_V(ZAHB)$  and 
metallicity was supported by  several HB stellar evolution models (Dorman 1992;  
Cassisi et al. 1999; VandenBerg et al. 2000; Catelan et al. 2004; Pietrinferni et al. 
2006) and pulsation models (Bono et al. 2007), and was confirmed by empirical HB data 
for 61 Galactic GCs (Ferraro et al. 1999).  

Our data do not show any significant deviation from a linear trend. 
The error bars are clearly larger than those of MW field or 
GC data, nevertheless the $M_V(HB)-[Fe/H]$ relation is well defined.

\section{Summary and Conclusions}\label{s:sum}

We have collected a homogeneous and uniform set of CMDs for 48 old GCs in M31 
obtained from $HST$ BVI data, 
to investigate the global characteristics of population II stars in this 
galaxy and compare them with those of the Milky Way. 

Of these CMDs, 35 were originally produced by our team during more than a decade
using basically the same criteria and procedures, 
and 13 were obtained by another group and re-derived by us to ensure the best 
possible homogeneity of the entire CMD set.

These CMDs were compared with template CMD ridge lines of selected 
Galactic GCs, and the best fit led to the simultaneous determination of 
reddening, metallicity, luminosity level of the horizontal branch $M_V(HB)$ 
and distance for each cluster. 

This set of parameters allowed us to derive the relation  
$$M_V(HB) = (0.25 \pm 0.02)[Fe/H] + (0.89 \pm 0.03)$$,  
where [Fe/H] is the cluster metallicity in the ZW84 scale.
 
By normalising this relation at the reference value of [Fe/H]=--1.5 to a similar relation 
using the apparent dereddened HB magnitude $V_0$(HB), we derived the distance modulus 
$(m-M)_0$(M31)=24.42$\pm$0.06 mag. 
This result agrees excellently with previous estimates from various 
distance indicators, and we consider it a robust and reliable estimate.
   
{\em This is the first determination of the distance to M31 based on the characteristics 
of its GC system calibrated on Galactic GC analogues.}  

The above relation also leads to a distance to the LMC of 18.54$\pm$0.07 mag, 
which excellently agrees with the value found by Benedict et al. (2011) using 
the $HST$ parallaxes of classical Cepheids and  RR Lyrae stars in the MW, as well as
other distance determinations.

\begin{acknowledgements}
We acknowledge the support by INAF through the PRIN-INAF 2009 grant CRA 
1.06.12.10 (PI: R. Gratton) and by ASI through contracts COFIS ASI-INAF 
I/016/07/0 and ASI-INAF I/009/10/0.
\end{acknowledgements}


\begin{thebibliography}{}

\bibitem[1996]{aja96} Ajhar, E. A., Grillmair, C. J., Lauer, T. R., Baum, W. A., Faber, S. M.,
  Holtzman, J. A., Lynds, C. R., \& O'Neil, E. J., Jr. 1996, \aj, 111, 1110
\bibitem[2000]{bar00} Barmby, P., Huchra, J.P., Brodie, J.P., Forbes, D.A., Schroder, L.L., \&
 Grillmair, C.J 2000, \aj, 119, 727
\bibitem[2011]{ben11} Benedict, G.F., McArthur, B.E., Feast, M.W. et al. 2011, 
     \aj, 142, 187       
\bibitem[2011]{bon11} Bonanos, A.Z., Castro, N., Macri, L.M. \&  Kudritzki, R.P 2011, 
     \apj, 729, L9
\bibitem[2007]{bono07} Bono, G., Caputo, F. and Di Criscienzo, M. 2007, \aap, 476, 779
\bibitem[2006]{bs06} Brodie, J.P., \& Strader, J. 2006, \araa, 44, 193 
\bibitem[2004]{bro04} Brown, T.M., Ferguson, H.C., Smith, E. et al. 2004, 
   \apj, 613, L125
\bibitem[2003]{cc03} Cacciari, C. \& Clementini, G. 2003, in {\it Stellar Candles for 
     the Extragalactic Distance Scale}, Ed.s D. Alloin and W. Gieren, Lecture Notes in 
     Physics, Vol. 635, p.105
\bibitem[2011]{cal11} Caldwell, N., Schiavon, R., Morrison, H., Rose, J.A. 
   and Harding, P. 2011 \aj, 141, 61 
\bibitem[1997]{calo97} Caloi, V., D'Antona, F. and Mazzitelli, I. 1997, \aap, 320, 823  
\bibitem[2011]{cap11} Caputo, F. 2012, \apss, in press  (Online First) 
\bibitem[2000]{cap00} Caputo, F., Castellani, V., Marconi, M. and Ripepi, V.  2000, 
     \mnras, 316, 819
\bibitem[[2009]{cg09} Carretta, E., Bragaglia, A., Gratton, R., D'Orazi, V. and 
   Lucatello, S. 2009, \aap, 508, 695 
\bibitem[1999]{cas99} Cassisi, S., Castellani, V., Degl'Innocenti, S., Salaris, M. 
     and Weiss, A.  1999, \aaps, 134, 103
\bibitem[1996]{cat96} Catelan, M. \& de Freitas Pacheco, J.A. 1996, \pasp, 108, 166 
\bibitem[2004]{cat04} Catelan, M., Pritzl, B.J. \& Smith, H.A. 2004, \apjs, 154, 633
\bibitem[1999]{chab99} Chaboyer, B.  1999,  Astrophysics \& Space Science, 237, 111
\bibitem[2003]{clem03} Clementini, G., Gratton, R., Bragaglia, A., Carretta, E., Di Fabrizio, L. 
     and Maio, M. 2003, \aj, 125, 1309 
\bibitem[2008]{clem09} Clementini, G. 2009, in {\it The Magellanic System: Stars, Gas, and Galaxies}, 
     IAU Symp. Vol. 256, p. 373 (arXiv:0906.1674)
\bibitem[2009]{col09} Colucci, J.E., Bernstein, R.A., Cameron, S., McWilliam, A. 
 \& Cohen, J.G. 2009, \apj, 704, 385 
\bibitem[2012]{con12} Contreras Ramos, R., Clementini, G., Federici, L. et al.   2012, 
in preparation 
\bibitem[2011]{dar11} Darling, J. 2011,  \apj, 732, L2  
\bibitem[2000]{dem00} Demarque, P., Zinn, R., Lee, Y-W. and Yi, S. 2000, \aj, 119, 1398
\bibitem[1992]{dor92} Dorman, B. 1992, \apjs, 81, 221 
\bibitem[2010]{dot10} Dotter, A., Sarajedini, A., Anderson, J. et al. 2010, \apj, 708, 698 
\bibitem[1962]{els} Eggen, O.J., Lynden-Bell, D. \& Sandage, A.R. 1962, 
      \apj, 136, 748 
\bibitem[2008]{fan08} Fan, Z., Ma, J., de Grijs, R. and  Zhou, X. 2008, 
   \mnras, 385, 1973 
\bibitem[1997]{fea97} Feast, M.W. 1997, \mnras, 284, 761
\bibitem[1992]{feba92} Feigelson, E.D., \& Babu, G.J. 1992, \apj, 397, 55
\bibitem[1998]{fern98} Fernley, J., Carney, B.W., Skillen, I., Cacciari, C. and Janes, K. 1998, 
     \mnras, 293, L61 
\bibitem[1999]{fer99} Ferraro, F.R., Messineo, M., Fusi Pecci, F. et al. 1999, \aj, 118, 1738
\bibitem[2003]{fit03} Fitzpatrick, E.L., Ribas, I., Guinan, E.F., Maloney, F.P. \& 
     Claret, A. 2003, \apj, 587, 685 
\bibitem[2011]{freed11}  Freedman, W.L., Madore, B.F., Scowcroft, V., et al. 2011, \aj, 142, 192 
\bibitem[2002]{fbh02} Freeman, K.C. \& Bland-Hawthorn, J. 2002, \araa, 40, 487 
\bibitem[1996]{fp96}  Fusi Pecci, F., Buonanno, R., Cacciari, C., et al. 1996, AJ, 112, 1461 [FFP96]
\bibitem[2004]{gal04} Galleti, S., Federici, L., Bellazzini, M., Fusi Pecci, 
     F. and Macrina, S. 2004, \aap, 416, 917 [RBCV4.0] 
\bibitem[2006]{gal06} Galleti, S., Federici, L., Bellazzini, M., Buzzoni, 
     A. and Fusi Pecci, F.  2006, \apj, 650, L107 [G06]
\bibitem[2009]{gal09} Galleti, S., Bellazzini, M., Buzzoni, A., Federici, L. and 
     Fusi Pecci, F. 2009, \aap,  508, 1285
\bibitem[2003]{gra03} Gratton, R.G., Bragaglia, A., Carretta, E.  et al. 2003, \aap, 408, 529  
\bibitem[2004]{gra04} Gratton, R.G., Bragaglia, A., Clementini, G.  et al. 2004, \aap, 421, 937   
\bibitem[2007]{han07} Hansen, B.M.S., Anderson, J., Brewer, J. et al. 2007, \apj, 671, 380   
\bibitem[1991]{har91} Harris, W.E. 1991 \araa, 29, 543 
\bibitem[1996]{har96} Harris, W.E. 1996, \aj, 112, 1487     
\bibitem[1995]{hol95} Holtzman, J.A., Burrows, C.J., Casertano, S. et al. 1995, 
     \pasp, 107, 1065 
\bibitem[2000]{jab00} Jablonka, P., Courbin, F., Meylan, G. et al. 2000, \aap, 359, 131 
\bibitem[1959]{kin59} Kinman, T.D. 1959,  \mnras, 119, 538  
\bibitem[1994]{lee94} Lee, Y.-W., Demarque, P., \& Zinn, R. 1994, \apj, 423, 248 
\bibitem[2006]{M06} Mackey, A.D., Huxor, A.P., Ferguson, A.M.N. et al. 2006,  \apj, 653, L105 [M06]
\bibitem[2007]{M07} Mackey, A.D., Huxor, A.P., Ferguson, A.M.N. et al. 2007, \apj, 655, L85  [M07]
\bibitem[2005]{mc05} Marconi, M. and Clementini, G. 2005, \aj, 129, 2257
\bibitem[1997a]{mcn97a} McNamara, D.H. 1997, \pasp, 109, 857
\bibitem[1999]{mcn99} McNamara, D.H. 1999, \pasp, 111, 489
\bibitem[2009]{mon09} Montalto, M., Seitz, S., Riffeser, A., Hopp, U., Lee, C.-H. 
     and Schoenrich, R.  2009, \aap, 507, 283
\bibitem[1959]{mor59} Morgan, W.W. 1959,  \aj, 64, 432
\bibitem[2009]{per09} Perina, S., Federici, L., Bellazzini, M. et al. 2009, 
   \aap, 507, 1375 [P09]
\bibitem[2010]{per11} Perina, S., Galleti, S., Fusi Pecci, F., Bellazzini, M., Federici, L., 
        \& Buzzoni, A. 2011, \aap, 531, 155  [P11] 
\bibitem[2006]{basti06} Pietrinferni, A., Cassisi, S., Salaris, M. and Castelli, F. 2006, 
     \apj, 642, 797
\bibitem[2009]{piet09} Pietrzy\'nski, G., Thompson, I.B., Graczyk, D. et al. 2009, \apj, 
     697, 862
\bibitem[2002]{pio02} Piotto, G., et al. 2002, \aap, 391, 945
\bibitem[1997]{reid97}  Reid, I.N. 1997, \aj, 114, 161 
\bibitem[2012]{rej12} Rejkuba, M. 2012, \apss, in press (Online First), arXiv:1201.3936
\bibitem[1996]{ren96} Renzini, A., Bragaglia, A., Ferraro, F.R. et al. 1996, \apj, 465, L23 
\bibitem[2005]{rich05} Rich, R.M., Corsi, C.E., Cacciari, C. et al. 2005, 
     \aj, 129, 2670 [R05]
\bibitem[2012]{rie12} Riess, A.G., Fliri, J., \& Valls-Gabaud, D. 2012, \apj, 745, 156
\bibitem[2000a]{ros00a} Rosenberg, A., Piotto, G., Saviane, I. \& Aparicio, A. 2000a, \aaps, 144, 5
\bibitem[2000b]{ros00b} Rosenberg, A., Aparicio, A., Saviane, I. \& Piotto, G. 2000b, \aaps, 145, 451
\bibitem[2012]{sal12} Salaris, M. 2012, \apss, in press (Online First) 
\bibitem[1993]{san93} Sandage, A. 1993, \aj, 106, 703 
\bibitem[1995]{sar95} Sarajedini, A., Lee, Y.-W. and Lee, D.H. 1995, \apj, 450, 712  
\bibitem[1998]{sch98} Schlegel, D. J., Finkbeiner, D. P., \& Davis, M. 1998, 
   \apj, 500, 525       
\bibitem[1978]{sz78} Searle, L. \& Zinn, R. 1978, \apj, 225, 357
\bibitem[2005]{sir05} Sirianni, M., Jee, M.J., Ben\'{i}tez, N. et al. 2005, \pasp, 117, 1049 
\bibitem[2001]{ste01} Stephens, A.W., Frogel, J.A., Freedman, W. et al. 2001, \aj, 121, 2597
\bibitem[2010]{tho10} Thompson, I.B., Kaluzny, J., Rucinski, S.M., Krzeminsky, W., Rych, W., Dotter, A.,
 \& Burley, G.S. 2010, \aj, 139, 329 
\bibitem[2000]{vdb00} VandenBerg, D.A., Swenson, F.J., Rogers, F.J., Iglesias, C.A. and 
     Alexander, D.R.  2000, \apj, 532, 430
\bibitem[2006]{vil06} Vilardell, F.,Ribas, I., Jordi, C. 2006, \aap, 459, 321 
\bibitem[2010]{vil10} Vilardell, F., Ribas, I., Jordi, C., Fitzpatrick, E. L. \& Guinan, E. F.
     2010, \aap, 509, 70 
\bibitem[2011]{wal11} Walker, A.R. 2011, \apss, in press (Online First), arXiv:1112.3171 
\bibitem[1984]{zw84} Zinn, R. and West, M.J. 1984, \apjs, 55, 45  [ZW84]
\bibitem[1985]{z85}  Zinn, R. 1985, \apj, 293, 424
  

%
%
%
%
%
%
\end{thebibliography}
\end{document}